\documentclass[10pt]{article}

\usepackage{latexsym}
\usepackage{amsfonts}
\usepackage{amsmath}
\usepackage{amsthm}
\usepackage{amssymb}
\usepackage{hyperref}

\def\d{\partial}

\newcommand{\dd}[2]{\frac{\d #1}{\d #2}}

\newcommand{\vdd}[2]{{\frac{\delta #1}{\delta #2}}}

\def\half{{\frac{1}{2}}}

\def\be{\begin{eqnarray}}
\def\ee{\end{eqnarray}}
\def\beann{\begin{eqnarray*}}
\def\eeann{\end{eqnarray*}}
\def\beq{\begin{equation}}
\def\eeq{\end{equation}}
\def\ba{\begin{array}}
\def\ea{\end{array}}
\def\ben{\begin{enumerate}}
\def\een{\end{enumerate}}
\def\bea{\begin{eqnarray}}
\def\eea{\end{eqnarray}}
\def\beann{\begin{eqnarray*}}
\def\eeann{\end{eqnarray*}}
\def\beq{\begin{equation}}
\def\eeq{\end{equation}}
\def\ba{\begin{array}}
\def\ea{\end{array}}
\def\ben{\begin{enumerate}}
\def\een{\end{enumerate}}

\def\5{\bar }
\def\6{\partial }
\def\7{\hat }
\def\4{\tilde }

\def\s0#1#2{\mbox{\small{$\frac{#1}{#2}$}}}

\newtheorem{corollary}{Corollary}

\newtheorem{theorem}{Theorem}

\def\qed{\hbox{${\vcenter{\vbox{
\hrule height 0.4pt\hbox{\vrule width 0.4pt height 6pt
\kern5pt\vrule width 0.4pt}\hrule height 0.4pt}}}$}}

\begin{document}


\begin{titlepage}
\begin{flushright}
ULB-TH/03-01\\
\end{flushright}

\begin{centering}

\vspace{0.5cm}

{\huge Boundary charges in gauge theories: \\using Stokes theorem
in the bulk} \\

\vspace{1.5cm}

{\large Glenn Barnich$^{*}$}

\vspace{.5cm}

Physique Th\'eorique et Math\'ematique,\\ Universit\'e Libre de
Bruxelles,\\
Campus Plaine C.P. 231, B--1050 Bruxelles, Belgium

\vspace{3cm}

\begin{abstract}
Boundary charges in gauge theories (like the ADM mass in general
relativity) can be understood as integrals of linear conserved n-2
forms of the free theory obtained by linearization around the
background. These forms are associated one-to-one to reducibility
parameters of this background (like the time-like Killing vector
of Minkowski space-time). In this paper, closed n-2 forms in the
full interacting theory are constructed in terms of a one
parameter family of solutions to the full equations of motion that
admits a reducibility parameter. These forms thus allow one to
apply Stokes theorem without bulk contributions and, provided
appropriate fall-off conditions are satisfied, they reduce
asymptotically near the boundary to the conserved n-2 forms of the
linearized theory. As an application, the first law of black hole
mechanics in asymptotically anti-de Sitter space-times is derived.
\end{abstract}

\end{centering}

\vfill

\footnotesize{$^*$Research Associate of the Belgium National Fund
for Scientific Research.}

\end{titlepage}

\section{Introduction}

Both Lagrangian \cite{Misner:1970aa,Abbott:1982ff,Abbott:1982jh}
and Hamiltonian \cite{ADM,Regge:1974zd} approaches involve in some
way the idea that it is the linearized theory around the
background that determines the asymptotically conserved $n-2$
forms used for the construction of boundary charges in gauge
theories in $n$ space-time dimensions. Recent results from
variational calculus \cite{Barnich:1995db} (see
\cite{Barnich:2000zw} for a review and also
\cite{Anderson:1996sc,Torre:1997cd}) corroborate this point of
view:
\begin{itemize}
\item when restricted to solutions of the equations of motion,
equivalence classes of closed, local, $n-2$ forms up to exact,
local, $n-2$ forms correspond one-to-one to non trivial
reducibility parameters; in this context, reducibility parameters
are possibly field dependent gauge parameters such that the
associated gauge transformations vanish on solutions of the
equations of motion; such parameters are trivial if they vanish
themselves on solutions of the equations of motion;
\item in standard interacting gauge theories like general
relativity or semi-simple Yang-Mills theories in space-time
dimensions strictly higher than 2, there are no non trivial
reducibility parameters and thus no non trivial conserved $n-2$
forms; in other words, every local $n-2$ form that is closed on
solutions of the equations of motion is given by the exterior
derivative of a local $n-3$ form on solutions of the equations of
motion;
\item in linear gauge theories however, reducibility parameters may
very well exist; for instance in general relativity linearized
around some background, particular reducibility parameters are
given by the Killing vectors of the background; furthermore, for
the flat background in space-time dimensions strictly higher than
2, they can be shown to be the only non trivial ones
\cite{Boulanger:2000rq}.
\end{itemize}

In \cite{Barnich:2001jy}, the one-to-one correspondence has been
extended to (suitable equivalence classes of) asymptotically
conserved $n-2$ forms on the one hand and asymptotic reducibility
parameters on the other hand. Furthermore, for given reducibility
parameters, the asymptotically conserved $n-2$ forms have been
explicitly constructed out of the linearized equations of motion
and of the gauge transformations evaluated at the background. That
the associated charges have all the standard properties like time
independence or independence of the form or position of the closed
$n-2$ dimensional hypersurface used in their definition is a
direct consequence of on-shell closure and of Stokes theorem. By
construction however, the $n-2$ forms are only closed near the
boundary, when evaluated for asymptotic solutions, i.e.,
deviations from the background that satisfy the boundary
conditions and the linearized field equations to leading order.
Hence, the application of Stokes theorem to relate the boundary
charges to integrals over surfaces deep in the bulk will in
general involve bulk contributions.

From the point of view of the full interacting theory, different
expressions for asymptotically conserved $n-2$ forms are
considered as equivalent to the linear $n-2$ forms discussed
above, if asymptotically near the boundary, all non linear terms
in the field deviations from the background vanish and if the
linear terms belong to the same equivalence class (and thus define
the same boundary charges) as the $n-2$ forms of the linearized
theory. This leaves of course a lot of freedom in the definition
of these forms, and allows one to show for instance that the
expressions derived in
\cite{LL,Misner:1970aa,Abbott:1982ff,Anderson:1996sc} for
energy-momentum and angular momentum in asymptotically flat
general relativity are all equivalent (see also
e.g.~\cite{Bak:1994gf} for a recent discussion).

Motivated by the work of Wald and Iyer
\cite{Wald:1993nt,Iyer:1994ys,Iyer:1995kg} (see also
e.g.\cite{Jacobson:1994vj}) on the formulation of the first law of
black hole mechanics in terms of Noether charge, we will construct
in this work $n-2$ forms of the full interacting theory that are
closed in a region of the bulk, provided that
\begin{itemize}

\item the $n-2$ form is constructed using a one parameter family
of solutions to the full equations of motion valid in the region
of the bulk where one wants to use Stokes theorem;

\item the exact reducibility parameters of the background are
simultaneously exact reducibility parameters of the one parameter
family of solutions to the full equations of motion.

\end{itemize}

As a result, for these one parameter family of solutions, the
boundary charges are related to the integrals of the $n-2$ forms
over surfaces deep in the bulk. Furthermore, if the Taylor
expansions in the parameter of the $n-2$ forms satisfy suitable
fall-off conditions near the boundary, these forms reduce to the
$n-2$ forms of the linearized theory discussed previously and thus
correctly describe the boundary charges.

In the next section, we  briefly review, in the context of the
linearized theory that is supposed to describe the full theory
asymptotically near the boundary, the expression for the linear
conserved $n-2$ forms associated to the reducibility parameters of
the background. Section 3 contains the main result on how the
conserved $n-2$ forms of the linearized theory should be modified
so that Stokes theorem can be used without bulk contributions. In
section 4, the theorem of section 3 is applied to the well known
cases of Yang-Mills theory and Einstein gravity with cosmological
constant. It is shown explicitly how the improved $n-2$ forms can
be used to express conservation of total energy. As an
application, a derivation of the first law of black hole mechanics
for asymptotically anti-de Sitter space-times is presented. In the
conclusion, the results that have been obtained are discussed from
the point of view of the original derivation of the first law and
comments on the relation to other approaches are made.

\section{Construction of the linear conserved n-2 forms of the free theory}

Let $R^i_\alpha[\phi]$ denote a generating set of gauge
transformations \cite{Henneaux:1992ig}, with associated gauge
symmetries $\delta_f\phi^i=R^i_\alpha(f^\alpha)$, where the
parameters $f^\alpha$ are local functions, i.e., they may depend
on $x^\mu$, the fields $\phi^i$ and a finite number of their
derivatives.

For example, in Einstein-Maxwell theory with cosmological constant
$\Lambda$, described by the action, \bea S=\int d^nx\
\frac{\sqrt{|g|}}{16\pi}[R-2\Lambda-F_{\mu\nu}F^{\mu\nu}], \eea
the fields $\phi^i$ correspond to $g_{\mu\nu},A_\mu$, the metric
and the electromagnetic gauge potentials respectively; the gauge
parameters $f^\alpha$ correspond to $\xi^\mu,\lambda$, the
parameters of an infinitesimal diffeomorphism and an infinitesimal
$U(1)$ gauge transformation respectively, while
$\delta_f\phi^i=R^i_\alpha(f^\alpha)$ corresponds to \bea
\delta_{\xi,\lambda}g_{\mu\nu}=L_\xi g_{\mu\nu},\\
\delta_{\xi,\lambda}A_\mu=L_\xi A_\mu+\partial_\mu \lambda,\eea
with $L_\xi$ denoting the Lie derivative.

For all $Q_id^nx$ with $Q_i$ local functions, we define the
current $n-1$ form $S^i_\alpha[\phi](Q_i,f^\alpha)=S^{i\mu}_\alpha
(Q_i,f^\alpha)(d^{n-1}x)_\mu$ through \bea
Q_iR^i_\alpha(f^\alpha)=f^\alpha R^{+i}_\alpha(Q_i)d^nx+d_H
S^i_\alpha(Q_i,f^\alpha).\label{defs} \eea In this equation,
$R^{+i}_\alpha$ denote the associated generating set of Noether
operators, obtained from the gauge symmetries by the integrations
by parts that move the derivatives from the gauge parameters to
the $Q_i$. The operator $d_H=dx^\mu\partial_\mu$, with
$\partial_\mu$ the total derivative with respect to $x^\mu$, is
the horizontal differential and
$(d^{n-p}x)_{\mu_1\dots\mu_p}=\frac{1}{p!(n-p)!}
\epsilon_{\mu_1\dots\mu_p}dx^{\mu_{p+1}}\dots dx^{\mu_n}$ with
$\epsilon_{0\dots n-1}=1$. For example, if \bea
Q_iR^i_\alpha(f^\alpha)=Q_iR^i_\alpha f^\alpha+Q_i R^{i\mu}_\alpha
\partial_\mu f^\alpha,\eea then \bea f^\alpha
R^{+i}_\alpha(Q_i)=f^\alpha R^i_\alpha Q_i
-f^\alpha\partial_\mu[R^{i\mu}_\alpha Q_i],\\S^{i\mu}_\alpha
(Q_i,f^\alpha)=Q_i R^{i\mu}_\alpha f^\alpha.\eea The defining
property of the Noether operators are the Noether identities \bea
R^{+i}_\alpha(\vdd{L}{\phi^i})=0.\label{ni}\eea  In the
Einstein-Maxwell example considered above, these identities
correspond to the contracted Bianchi identities $D_\mu
(G^{\mu\nu}+\Lambda g^{\mu\nu})=0$ and $D_\mu D_\nu F^{\nu\mu}=0$.

When the $Q_i$ are replaced with the left hand side of the
Euler-Lagrange equations of motion in \eqref{defs}, one gets,
because of the Noether identities \eqref{ni}, \bea
\vdd{L}{\phi^i}R^i_\alpha(f^{\alpha})=d_H S_{f},\ S_{f}\equiv
S^i_\alpha(\vdd{L}{\phi^i},f^{\alpha}). \label{full}\eea In other
words, $S_{f}$ is a weakly vanishing representative for the
Noether current $n-1$ form associated to the gauge symmetry
$\delta_f\phi^i=R^i_\alpha(f^\alpha)$.

In the Einstein-Maxwell example, $S_f$ is explicitly given by \bea
S_{\xi,\lambda}=\frac{\sqrt{|g|}}{16\pi}[(-G^{\mu\nu}-\Lambda
g^{\mu\nu}+8\pi T^{\mu\nu}_{em}+4D_\nu F^{\nu\mu}(\xi^\rho
A_\rho+\lambda)](d^{n-1}x)_\mu,\eea where the electromagnetic
energy-momentum tensor is \bea
T^{\mu\nu}_{em}=\frac{2}{\sqrt{|g|}}\vdd{L^{em}}{g_{\mu\nu}}=\frac{1}{4\pi}[
F^{\mu\alpha}{F^\nu}_\alpha-\frac{1}{4}
g^{\mu\nu}F^{\alpha\beta}F_{\alpha\beta}].\eea

Let ${\bar\phi}^i(x)$ be a background solution to the
Euler-Lagrange equations of motion and
$\phi^i={\bar\phi}^i(x)+\varphi^i$. We denote by $\frac{\delta
L^{\rm free}}{\delta\varphi^i}[\varphi;\bar\phi]$ the equations of
motion linearized around ${\bar\phi}^i(x)$, obtained from the
quadratic piece of the Lagrangian in an expansion according to
$\varphi$. If $f^{0\alpha}=f^\alpha[x,\bar\phi]$, the
linearization around ${\bar\phi}^i(x)$ of \eqref{full} gives \bea
\vdd{L^{\rm free} }{\varphi^i}R^{0i}_\alpha(f^{0\alpha})=d_H
s_{f^0},\ s_{f^0}[\varphi;\bar\phi]= S^{0i}_\alpha(\vdd{L^{\rm
free} }{\varphi^i},f^{0\alpha}), \label{lin}\eea where
$S^{0i}_\alpha=S^i_\alpha[\bar\phi(x)]$. For field independent
parameters\footnote{The general case of field dependent gauge
parameters $\tilde f^\alpha$ that satisfy
$R^{0i}_\alpha(f^\alpha)\approx^{free} 0$, where $\approx^{free}$
means an equality that holds on solutions of the equations of
motion of the linearized theory, is treated in section 3 of
\cite{Barnich:2001jy}. It can be shown that, when evaluated on
solutions of the linearized theory, the expression of the n-2
forms constructed below is still valid in the more general case.}
$\tilde f^\alpha$ that satisfy \bea R^{0i}_\alpha(\tilde
f^\alpha)= 0\label{linred},\eea equation \eqref{lin} implies that
\bea d_Hs_{\tilde f}=0\label{linclos}.\eea In the Einstein-Maxwell
example, solutions to \eqref{linred} are given by the Killing
vectors of the background metric, $L_{\tilde \xi} \bar
g_{\mu\nu}=0$, which satisfy in addition $L_{\tilde \xi} \bar
A_\mu+\partial_\mu\tilde\lambda=0$ for some gauge parameter
$\tilde\lambda$.

In the case of trivial topology, equation \eqref{linclos} implies
that $s_{\tilde f}=d_H \tilde k_{\tilde f}$, for some $n-2$ form
$\tilde k_{\tilde f}[\varphi;\bar\phi]$. Provided the equations of
motion $\frac{\delta L}{\delta\phi^i}$ and the gauge
transformations $R^i_\alpha(f^\alpha)$ are local, in the sense
that they depend only on a finite number of derivatives of the
fields, the $n-2$ form $\tilde k_{\tilde f}[\varphi;\bar\phi]$ can
be constructed to be local as well, in the sense that it depends
(linearly) on the $\varphi^i$ and a finite number of their
derivatives, and on a finite number of derivatives of the
background $\bar\phi^i$. How to explicitly construct $\tilde
k_{\tilde f}$ out of $s_{\tilde f}$ has been explained in many
references,
see e.g.~\cite{Vinogradov:1977,Takens:1979aa,Tulczyjew:1980aa,Anderson:1980aa,%
DeWilde:1981aa,Tsujishita:1982aa,Brandt:1990gy,Dubois-Violette:1991is,%
Dickey:1992aa,Dragon:1996md}. In the general relativity literature
for instance, an algorithm has been given in \cite{Wald:1990aa}.
In fact, the whole theory of "black hole entropy from Noether
charge" \cite{Wald:1993nt,Iyer:1994ys,Jacobson:1994vj,Iyer:1995kg}
relies crucially on this algorithm. Whereas in these references,
the algorithm is used in terms of arbitrary gauge parameters, we
will use the explicit formula for the "contracting homotopy"
involving higher order Euler operators due to Anderson
\cite{Olver:1993,Andersonbook}, in terms of the fields $\varphi^i$
of the linearized theory below, and in terms of the fields
$\phi^i$ of the full theory in the next section. Indeed, in the
case of trivial topology, for local forms $\omega^p$ of degree $p$
strictly lower than $n$ that vanish when the $\varphi^i$ and their
derivatives are set to zero, there exists an operator
$\rho_{H,\varphi}$ such that \bea
d_H(\rho^p_{H,\varphi}\omega^p)+\rho^{p+1}_{H,\varphi}(d_H\omega^p)=
\omega^p\label{inv}. \eea It follows that the $n-2$ form of the
linearized theory defined by \bea\tilde k_{\tilde
f}=\rho^{n-1}_{H,\varphi}s_{\tilde f},\label{linform}\eea is
closed when the linearized equations of motion hold: \bea d_H
\tilde k_{\tilde f}=s_{\tilde f}\approx^{\rm free}0.\eea The
explicit expression for $\tilde k_{\tilde f}$ is \bea \tilde
k_{\tilde
f}[\varphi;\bar\phi]=\sum_{k=0}\frac{k+1}{k+2}\partial_{\mu_1}\dots
\partial_{\mu_k}\left[\varphi^j
\vdd{}{\varphi^j_{\mu_1\dots\mu_k\rho}}\dd{s_{\tilde
f}}{dx^\rho}\right],\label{expllinex}\eea with the understanding
that the first term in the sum does not contain an index $\mu$ and
no total derivative. Furthermore, the notation $\dd{}{dx^\rho}$
stands for the contraction of a form with the vector
$\dd{}{x^\rho}$, while the higher order Euler operators
$\vdd{}{\varphi^j_{\mu_1\dots\mu_k\rho}}$ are constructed out of
the symmetrized partial derivatives
$\dd{{}^S}{\varphi_{\mu_1\dots\mu_k}}$. The detailed combinatorial
factors involved in the definitions of these operators can be
found in \cite{Andersonbook} or in appendix A of
\cite{Barnich:2001jy}. Note however that the definition of $\tilde
k_{\tilde f}[\varphi;\bar\phi]$ used here differs by an overall
minus sign from the one used in \cite{Barnich:2001jy}.

The $n-2$ forms $\tilde k_{\tilde f}[\varphi;\bar\phi]$ have been
explicitly computed in \cite{Barnich:2001jy} for Yang-Mills theory
and Einstein gravity with cosmological constant and the results of
\cite{Abbott:1982ff,Abbott:1982jh} and \cite{Anderson:1996sc} have
been recovered. From the point of view of
\cite{Abbott:1982ff,Abbott:1982jh} the contracting homotopy
$\rho^p_{H,\varphi}$ provides a systematic way of converting
"volume integrals to surface integrals".

In the next section, we construct an $n-2$ form in the full theory
that, under suitable assumptions, is closed in the bulk and
reduces asymptotically to the $n-2$ form \eqref{expllinex} of the
linearized theory if appropriate fall-off conditions are
satisfied.

\section{Construction of the closed n-2 forms of the full theory}

The most general form of the homotopy formula for $d_H$ (given in
\cite{Andersonbook}, chapter 4, pages 119-122) allows one to
interpolate between a form evaluated at two different field
configurations by using a (not necessarily straight) path that
connects these configurations: if $\phi^i_s(x)$, $s\in [0,1]$ is a
one parameter family of field configurations and
$\varphi^i_s(x)=\frac{d\phi^i_s(x)}{ds}$, one can show that \bea
S_{f}[\phi_1(x)]-S_{f}[\phi_0(x)]&=& \int_0^1 ds
\frac{d}{ds}\left( S_{f}[\phi_s]\right)\nonumber\\&=&\int_0^1 ds
\left(\sum_{k=0}\partial_{\mu_1}\dots
\partial_{\mu_k}\varphi^i_s
\left[\frac{\partial^S}{\partial\phi^i_{\mu_1\dots\mu_k}}S_{ f}
\right][\phi_s]\right)\nonumber\\&= &d_H \rho_H^{n-1} S_{ f}
+\rho^n_H d_H S_{f}, \eea where for a $p$ form $\omega^p[\phi]$,
\bea \rho^p_H \omega^p=\int_0^1 ds\
I^p_{\varphi_s(x)}(\omega^p)[\phi_s(x)],\eea with \bea
I^p_{\varphi_s}(\omega^p)=\sum_{k=0}\frac{k+1}{n-p+k+1}
\6_{\mu_1}\dots\6_{\mu_k}
\left[\varphi^i_s(x)\vdd{}{\phi^i_{\mu_1\dots\mu_k\rho}}
\dd{\omega^p}{dx^\rho}\right]. \eea If we define \bea K_{
f}=\rho^{n-1}_{H}S_{f},\label{n-2form}\eea it follows using
\eqref{full} that \bea d_H K_{
f}=-\rho^n_{H}\left(\vdd{L}{\phi^i}R^i_\alpha(
f^\alpha)d^nx\right)+S_{f}[\phi_1(x)]- S_{f}
[\phi_0(x)],\label{23} \eea for an arbitrary configuration
$\phi^i_s(x)$. Using the explicit expression of $\rho^n_{H}$, one
can then show:
\begin{theorem}
For a given solution $\phi^i(x)$ of the full equations of motion,
the $n-2$ form $K_{\tilde f}$, defined in \eqref{n-2form}, is
closed in the bulk, \bea d_H K_{\tilde f}=0, \eea provided
\begin{itemize}
\item a one parameter family $\phi^i_s(x)$ of solutions to the full
equations of motion interpolating between $\phi^i(x)$ and the
background solution $\bar\phi^i(x)$ is used, \bea
\vdd{L}{\phi^i}[\phi_s(x)]=0\label{solfull}, \forall s\in
[0,1],\eea

\item the gauge parameters $f^\alpha=\tilde f^\alpha$ are
reducibility parameters for this interpolating solution, \bea
R^i_\alpha[\phi_s(x)](\tilde f^\alpha)=0, \forall s\in
[0,1].\label{redfull} \eea
\end{itemize}
\end{theorem}
The proof of the theorem is given in the appendix.

The $n-2$ form $K_{\tilde f}$ depends in general on the path
$\gamma$ in the space of solutions chosen to interpolate between
$\bar\phi(x)$ and $\phi(x)$, but not on the parameterization
chosen for this path\footnote{The author wants to thank J.~Zanelli
for drawing his attention to this point.}. This justifies the
following notation: \bea K_{\tilde f}=\int_{\gamma}
I^{n-1}_{d_V\phi} (S_{\tilde f})[\phi], \eea where $d_V\phi^i$
stands for an infinitesimal variation of the fields. More
precisely, $K_{\tilde f}$ is the integral along $\gamma$ of the
vertical $1$ form and horizontal $n-2$ form in the variational
bicomplex \cite{Olver:1993,Andersonbook}.

Given appropriate fall-off conditions and an analytic expansion in
$s$, the $n-2$ form $K_{\tilde f}$ coincides asymptotically near
the boundary with the linear $n-2$ form $\tilde k_{\tilde f}$
discussed in the previous section: indeed, if
$\phi_s(x)={\bar\phi}^i(x)+s\varphi^i(x) +s^2\varphi^i_2(x)+\dots$
and the fall-off conditions are such that, in an expansion
according to $s$, only the term independent of $s$ in
$I_{\varphi_s(x)}(S_{\tilde f})[\phi_s(x)]$ contributes because
all the other terms fall off too fast near the boundary, we have
\bea K_{\tilde f}\longrightarrow I_{\varphi(x)}(S_{\tilde
f})[\bar\phi(x)] \eea In the appendix, it is shown that this
expression agrees with expression \eqref{expllinex} of the
linearized theory.

This reasoning can also be turned around to see how $K_{\tilde f}$
can be constructed from $\tilde k_{\tilde f}[\varphi;\bar\phi]$ of
the linearized theory : because $s_{\tilde f}=(d_V S_{\tilde
f})[\varphi;\bar\phi]$, where $d_V$ denotes a variation of the
fields $\phi^i$ and their derivatives and the argument
$[\varphi;\bar\phi]$ means that the variations of the fields are
substituted by $\varphi^i$ and the fields by the background
solution $\bar\phi^i(x)$, $I^{n-1}_{\varphi_s(x)} (S_{\tilde
f})[\phi_s(x)]$ is given by the right hand side of
\eqref{expllinex}, where $\varphi^i(x),{\bar\phi}^i(x)$ are
replaced by $\varphi^i_s(x),\phi^i_s(x)$. We have thus shown the
following corollary to theorem 1:
\begin{corollary}
The closed forms $K_{\tilde f}$ associated to a one parameter
family of solutions $\phi^i_s(x)$ with reducibility parameters
$\tilde f^\alpha$ can be obtained from the conserved $n-2$ forms
$\tilde k_{\tilde f}[\varphi;\bar\phi]$ of the linearized theory
defined in \eqref{expllinex} by substituting the background for
the one parameter family of solutions $\phi^s(x)$, by substituting
the field deviations $\varphi$ by
$\varphi^s(x)=\frac{d\phi^s(x)}{ds}$ and by integrating over the
parameter: \bea K_{\tilde f}=\int_0^1ds\ \tilde k_{\tilde
f}[\varphi^s;\phi^s].\label{fullfromlin}\eea
\end{corollary}
Again, if $\gamma$ is the path in the space of solutions
interpolating between $\bar\phi^i(x)$ and $\phi^i(x)$,
reparameterization invariance allows one to write \bea K_{\tilde
f}=\int_\gamma\ \tilde k_{\tilde f}[d_V\phi;\phi]. \eea

Consider now the path $\gamma +\delta\gamma$ where $\delta\gamma$
is the ``straight path" between the solution $\phi^i(x)$ and the
infinitesimally close solution $\phi^i(x)+\delta\phi^i(x)$. If we
define $\delta K_{\tilde f}$ to be the variation of $K_{\tilde f}$
when evaluated on $\gamma +\delta\gamma$ and on $\gamma$, it
follows that \bea \delta K_{\tilde
f}\equiv\int_{\gamma+\delta\gamma}\ \tilde k_{\tilde
f}[d_V\phi;\phi]-\int_{\gamma}\ \tilde k_{\tilde f}[d_V\phi;\phi]=
\tilde k_{\tilde f}[\delta\phi(x);\phi(x)].\label{29} \eea
\section{Standard applications}

The $n-2$ forms of the linearized theory given by
\eqref{expllinex} have been explicitly computed, up to a
conventional overall sign difference, in \cite{Barnich:2001jy}
section 6 for Yang-Mills theory and general relativity. The
corresponding $n-2$ forms in the full theory constructed using
\eqref{fullfromlin} are briefly discussed in the next two
subsections.

\subsection{Yang-Mills theory}

In the Yang-Mills case, the $n-2$ forms of the linearized theory
agree with the ones found in \cite{Abbott:1982jh} if $\bar D_\mu
\tilde f=0$ is taken into account. Here $\bar D_\mu$ is the
background covariant derivative and $\tilde f=\tilde f^a T_a$ are
non abelian gauge parameters. In the full non linear theory,
application of \eqref{fullfromlin} then gives \bea K_{\tilde
f}&=&(d^{n-2}x)_{\mu\nu}\int^1_0 ds\ {\rm
Tr}(\tilde f f^{\mu\nu}_s ) \nonumber\\
&=&(d^{n-2}x)_{\mu\nu}{\rm Tr}\left(\tilde f(F^{\mu\nu}(x)-\bar
F^{\mu\nu}(x))\right),\eea where \bea D_\mu[A_s(x)]\tilde f=0\eea
has been taken into account and
$f^{\mu\nu}_s(x)=D^\mu[A_s(x)]a^\nu_s(x)-D^\nu[A_s(x)]a^\mu_s(x)$
with $a^s_\mu=\frac{d A^s_\mu}{ds}$. Equivalently, \bea K_{\tilde
f} &=& (d^{n-2}x)_{\mu\nu}\int^1_0 ds\ \left(\6^\mu {\rm
Tr}(\tilde f a^\nu_s)-\6^\nu {\rm Tr}(\tilde f a^\mu_s)\right)
\nonumber\\&=& 2(d^{n-2}x)_{\mu\nu}\6^\mu {\rm Tr}\left(\tilde f
(A^\nu(x)-\bar A^\nu(x))\right).\eea In this case, the result does
not depend on the path $A^s_{\mu}(x)$ in the space of solutions
but only on the end points $\bar A_\mu(x)$ and $A_\mu(x)$.

\subsection{General relativity with cosmological constant}

In the case of gravity with Lagrangian
\[L=\frac{1}{16\pi}\sqrt{-g}(R-2\Lambda),\]
direct application of \eqref{expllinex} gives $n-2$ forms in the
linearized theory that agree with those of \cite{Abbott:1982ff}.
According to \eqref{fullfromlin}, the $n-2$ forms in the non
linear theory are obtained by replacing the background $\bar
g_{\mu\nu}$ by a one parameter family of solutions $g_{\mu\nu}^s$
and the metric deviations $h_{\mu\nu}$ by $h^s_{\mu\nu}=\frac{d
g^s_{\mu\nu}}{ds}$ and integration the resulting expression over
$s$. Dropping for notational simplicity the $s$ dependence, one
finds \bea K_{\tilde
\xi}=\frac{1}{16\pi}(d^{n-2}x)_{\mu\nu}\int^1_0 ds\ \sqrt{-
g}\Big( \tilde\xi^\nu D^\mu h+\tilde \xi^\mu D_\sigma
h^{\sigma\mu}+\tilde \xi_\sigma D^\nu
h^{\sigma\mu}\nonumber\\+\half hD^\nu\tilde\xi^\mu+\half
h^{\mu\sigma}D_\sigma \tilde\xi^\nu+\half h^{\nu\sigma} D^\mu
\tilde\xi_\sigma -(\mu\longleftrightarrow \nu) \Big)
\\
=-\frac{1}{16\pi}(d^{n-2}x)_{\mu\nu}\int^1_0 ds\ \sqrt{-
g}\Big(\tilde \xi_\rho D_\sigma H^{\rho\sigma\mu\nu} +\frac 12
H^{\rho\sigma\mu\nu} \partial_\rho\tilde \xi_\sigma \Big),
\label{gsuperpot3} \eea where $H^{\rho\sigma\mu\nu}[h;g]$ has the
symmetries of the Riemann tensor: \bea
H^{\mu\alpha\nu\beta}[h;g]&=& -\7h^{\alpha\beta}g^{\mu\nu}
-\7h^{\mu\nu}g^{\alpha\beta} +\7h^{\alpha\nu}g^{\mu\beta}
+\7h^{\mu\beta}g^{\alpha\nu},
\label{Hdef}\\
\7h_{\mu\nu}&=& h_{\mu\nu}-\frac{1}{2} g_{\mu\nu} h. \label{hath}
\eea In these expressions, it is understood that $g_{\alpha\beta}=
g_{\alpha\beta}^s(x)$ is the metric used to define the covariant
derivative and, together with its inverse, to lower and raise the
indices, with $h=h^\mu_\mu$. In \cite{Barnich:2001jy}, it has been
pointed out that using the Killing equation $\bar
D_\mu\tilde\xi_\mu+\bar D_\nu\tilde\xi_\nu=0$, the corresponding
expression in the linearized theory agrees with the simplified
expression given in equation (11) of \cite{Anderson:1996sc}.
Accordingly, in the full theory, using the Killing equation \bea
D_\mu\tilde \xi_\nu+D_\nu\tilde \xi_\mu=0,\eea an equivalent
expression is \bea K_{\tilde \xi}=
\frac{1}{16\pi}(d^{n-2}x)_{\mu\nu}\int^1_0 ds\ \sqrt{-g}\Big(
h^{\mu\sigma}D_{\sigma}\tilde\xi^\nu-\tilde \xi_\sigma D^\mu
h^{\nu\sigma}- \half h D^\mu\tilde \xi^\nu \nonumber\\+\tilde
\xi^\mu (D_\sigma h^{\nu\sigma}-D^\nu h)-(\mu\longleftrightarrow
\nu)\Big).\label{anderson} \eea As pointed out in
\cite{Anderson:1996sc}, their expression (11) in the linearized
theory agrees with $d_V Q-\tilde \xi\cdot \Theta$, where $d_V$
denotes a field variation and $Q$ and $\Theta$ are defined in
equations (61) respectively (63) of \cite{Iyer:1994ys}. Using this
observation in the context of the full theory, one finds that
$K_{\tilde \xi}$ can be decomposed into a piece involving the
Komar integrands depending only on the end points of the path and
a path dependent piece that is proportional to the
undifferentiated Killing vector: \bea K_{\tilde \xi}=\int^1_0ds\
\Big(-\frac{1}{16\pi}\frac{d}{ds}K^K_{\tilde
\xi}[g_s(x)]-\tilde\xi^\nu
\frac{d}{dx^\nu}\Theta[g_s(x),h_s(x)]\Big)\nonumber\\
=-\frac{1}{16\pi}K^K_{\tilde \xi}[g(x)]+\frac{1}{16\pi}K^K_{\tilde
\xi}[\bar g(x)]-\int^1_0ds\ \tilde\xi^\nu \frac{\partial}{\partial
dx^\nu}\Theta[g_s(x),h_s(x)],\label{intform}\eea where \bea
K^K_{\tilde\xi}[g]=(d^{n-2}x)_{\mu\nu}
\sqrt{-g}\Big(D^\mu\tilde\xi^\nu-(\mu\leftrightarrow \nu)\Big)
\eea is the Komar integrand and  \bea
\Theta[g,h]=\frac{1}{16\pi}(d^{n-1}x)_\mu\sqrt{-g}\Big(D_\sigma
h^{\mu\sigma}-D^\mu h\Big).\eea

\subsection{The first law of black hole mechanics}

As an application, we consider $4$ dimensional asymptotically
anti-de Sitter, stationary and axisymmetric black hole space
times, such as for example the (uncharged) Kerr anti-de Sitter
black holes \cite{Carter:1968ks} (see also e.g.
\cite{Caldarelli:1999xj,Silva:2002jq} for recent discussions of
the first law in this context). In this particular case, the path
that interpolates between anti-de Sitter space $\bar g_{\mu\nu}$
and a given Kerr anti-de Sitter black hole
$g_{\mu\nu}(x)=g^1_{\mu\nu}(x)$ with fixed $M$ and $a$ can for
instance be chosen to be $g^s_{\mu\nu}(x)=g^{\mu\nu}(sM,sa)$.
(Note that this interpolation only needs to be valid in the region
of space-time where one wants to apply Stokes theorem.) The total
energy difference ${\cal E}$ between the background and the given
solution is \bea {\cal E}=\oint_{S^\infty} K_{k},\eea where
$S^\infty$ is the 2 sphere at infinity given in Boyer-Lindquist
type coordinates by $t=t_0, r=R\rightarrow \infty$, with $t_0,R$
constant, and the stationary Killing vector field $k$ in these
coordinates can (for instance) be chosen to be
$k=\frac{\partial}{\partial t}$.

As a consequence of Stokes theorem and the fact that $dK_{k}=0$,
conservation of total energy reduces to \bea {\cal E}=\oint_{S}
K_{k},\eea where $S$ is another closed $2$ dimensional surface
such that $S^\infty$ and $S$ are the boundaries of some $3$
dimensional volume $\Sigma$. In particular, time independence of
${\cal E}$ follows by choosing $S$ to be the $2$ sphere at
infinity for some other time $t=t_1$.

Let $S=H$ be the intersection of $\Sigma$ with the event horizon
$\cal H^+$, $k=\xi-\Omega_H m$, where $m$ is the axial Killing
vector field given in Boyer-Lindquist type coordinates by
$\frac{\partial}{\partial\phi}$, $\xi$ is the null generator of
the horizon associated to the solution
$g_{\mu\nu}(x)=g^1_{\mu\nu}(x)$ and $\Omega_H$ its constant
angular velocity. Using expression \eqref{intform} and the fact
that $m$ is tangent to $S^\infty$, a Smarr type formula can be
obtained as follows: \bea {\cal
E}&=&\oint_{S^\infty}[K_{\xi}-\Omega_HK_m]\nonumber\\&=&\oint_{H}K_\xi+\Omega_H
{\cal J}
\nonumber\\&=&\frac{\kappa}{8\pi}A_H+\frac{1}{16\pi}\oint_H
K^K_\xi[\bar g(x)]-\oint_H\int_0^1ds\ \xi^\nu
\frac{\partial}{\partial dx^\nu}\Theta+\Omega_H {\cal J},\eea with
$A_H$ the area of the horizon, $\kappa$ its surface gravity and
\bea {\cal
J}=-\oint_{S^{\infty}}K_m=\frac{1}{16\pi}\oint_{S^{\infty}}\Big(K^K_m[g(x)]-
K^K_m[\bar g(x)]\Big) \eea the total angular momentum.

In order to derive the first law of black hole mechanics in this
context, we will follow closely the reasoning of
\cite{Wald:1993nt,Iyer:1994ys} in the asymptotically flat case.
The same steps as in the previous paragraph will now be applied to
$\delta K_{k}$ given in \eqref{29}, with $H$ chosen to be the
bifurcation surface $B$ of the Killing horizon of the solution
$g_{\mu\nu}(x)=g^1_{\mu\nu}(x)$. If \bea \delta {\cal
J}&=&-\oint_{S^\infty} \tilde k_m[\delta
g(x);g(x)]\nonumber\\&=&\frac{1}{16\pi}\oint_{S^{\infty}}\delta
K^K_m
\nonumber\\&=&\frac{1}{16\pi}\oint_{S^{\infty}}(d^{n-2}x)_{\mu\nu}
\sqrt{-g}\Big( \tilde \xi_\sigma D^\mu \delta g^{\nu\sigma}-\delta
g^{\mu\sigma}D_{\sigma}\tilde\xi^\nu+ \nonumber\\& &\ \ \ \ \ \ \
\ \ \ \ \ \ \ \ \ \ \ \ \ \ \ \ \ \ \ \ \ \ \ \ \ \ \ \ \ +\half
\delta g D^\mu\tilde \xi^\nu-(\mu\leftrightarrow\nu)\Big),\eea
with $\delta g=\delta g^\mu_\mu$, we get \bea \delta {\cal
E}&\equiv& \oint_{S^\infty}\ \tilde k_k[\delta
g(x);g(x)]\nonumber\\&=&\oint_{S^\infty}\ \Big(\tilde k_\xi[\delta
g(x);g(x)]-\Omega_H\tilde k_m[\delta
g(x);g(x)]\Big)\nonumber\\&=&\oint_{B}\tilde k_\xi[\delta
g(x);g(x)]+\Omega_H\delta{\cal
J}\nonumber\\&=&\frac{\kappa}{8\pi}\delta A_H+\Omega_H\delta{\cal
J}.\eea The last line follows if one can show that
$\oint_{B}\tilde k_\xi[\delta g(x);g(x)]=\frac{\kappa}{8\pi}\delta
A_H$. Because $\xi$ vanishes on the bifurcation surface $B$, one
sees by comparing expressions \eqref{anderson} and \eqref{intform}
(without integral over $s$ and $h_{\mu\nu}$ replaced by $\delta
g_{\mu\nu}$) that only the variation of the Komar integrand can
contribute, $\oint_{B}\tilde k_\xi[\delta
g(x);g(x)]=-\frac{1}{16\pi}\oint_{B}\delta K^K_\xi$. That this
last expression reduces to $\frac{\kappa}{8\pi}\delta A_H$ can be
shown by following for instance the proof of theorem 6.1 of
\cite{Iyer:1994ys}.

\section{Conclusion}

In standard interacting gauge theories like semi-simple Yang-Mills
theory or general relativity in space-time dimensions strictly
higher than 2, all local $n-2$ forms that are closed on-shell are
trivial in the sense that they are given, on-shell, by the
exterior derivative of local $n-3$ forms. In the linearized theory
around the background, however, non trivial on-shell closed $n-2$
forms do exist and they are in one-to-one correspondence with non
trivial reducibility parameters of the background.

In the case where there exists a path in the space of solutions
that connects the background to the solution of interest and that
admits reducibility parameters, we have constructed in this paper
closed $n-2$ forms of the full interacting theory, that under
suitable assumptions, reduce asymptotically to the conserved $n-2$
forms of the linearized theory used in the definition of the
boundary charges. As a consequence, Stokes theorem can be used to
easily relate the boundary charges to the integral of these $n-2$
forms on surfaces deep in the bulk, as needs to be done for
instance in a derivation of the first law of black hole mechanics.

The standard derivation of the first law is based on Komar
integrals \cite{Bardeen:1973gs,Carter:1972} (see e.g.
\cite{Townsend:1997ku} for a review). Komar integrals are very
useful in this context because they allow one to directly relate
integrals defined over the 2 sphere at infinity to integrals
defined over the horizon, the bulk contribution that arises in a
direct application of Stokes theorem being easily expressible in
terms of the energy-momentum tensor. However, the Komar integrals
do not provide a complete theory for boundary charges: their
normalization has to be fixed by comparing to ADM type expressions
and their validity is restricted to the asymptotically flat case.

Systematic Lagrangian approaches to constructing the charges of
interacting gauge theories such as general relativity are based on
the linearized theory around the background
\cite{Abbott:1982ff,Anderson:1996sc,Barnich:2001jy}. When applying
Stokes theorem directly to the corresponding n-2 forms in order to
reach surfaces deep in the bulk such as the black hole horizon,
one has to take into account the complicated bulk contribution of
the non linear part of the field equations that have been shuffled
into the right hand side of the equations of motion in the form of
an effective energy momentum tensor.

In this paper, we have shown how complicated bulk contributions
can be avoided rather easily by a straightforward improvement of
the n-2 forms constructed in the linearized theory. Furthermore,
because the n-2 forms constructed here are directly related to the
boundary charges for non flat backgrounds, they allow one to
generalize the approach of
\cite{Wald:1993nt,Iyer:1994ys,Jacobson:1994vj,Iyer:1995kg} to the
first law of black hole mechanics to such more general
backgrounds.

In future work, we plan to study general conditions under which
the $n-2$ forms are path independent and apply the formalism to
more complicated theories, for instance the higher curvature
gravity theories considered in
\cite{Jacobson:1993xs,Anderson:1996sc,Deser:2002rt,Deser:2002jk},
or more exotic black hole solutions involving scalar fields like
those discussed in \cite{Henneaux:2002wm,Martinez:2002ru}.

\section*{Acknowledgements}

The author wants to thank R.~Aros, M. Ba\~nados, X.~Bekaert,
F.~Brandt, C.~Chryssomalakos, V.~Frolov, M.~Henneaux, C.~Mart\'{\i}nez,
R.~Olea, R.~Troncoso and J.~Zanelli for useful discussions. This
work is supported in part by the ``Actions de Recherche
Concert\'ees'' of the ``Direction de la Recherche
Scientifique-Communaut\'e Fran\c caise de Belgique, by a ``P\^ole
d'Attraction Interuniversitaire'' (Belgium), by IISN-Belgium
(convention 4.4505.86), by Proyectos FONDECYT 1970151 and 7960001
(Chile) and by the European Commission RTN programme HPRN-CT00131,
in which the author is associated to K.~U.~Leuven.

\appendix
\section*{Appendix}
\setcounter{equation}{0}
\def\theequation{A.\arabic{equation}}
In the first part of this appendix, we prove theorem 1 and derive
sufficient conditions that guarantee that $K_{ f}$ defined by
\eqref{n-2form} is closed in the bulk. If
$\phi^i_0(x)\equiv{\bar\phi}^i(x)$ and
$\phi^i_1(x)\equiv\phi^i(x)$ are solutions to the full equations
of motion, the last two terms on the right hand side of \eqref{23}
vanish. Taking into account the explicit expression for the higher
order Euler operators, the first term on the right hand side of
\eqref{23} is given by \bea
\rho^n_{H}\left(\vdd{L}{\phi^j}R^j_\alpha(
f^\alpha)d^nx\right)=\nonumber\\\int^1_0 ds\
\6_{(\mu)}\Big[\varphi^i_s(x)
\left(\begin{array}{c}|\mu|+1+|\nu|\\
|\mu|+1\end{array}\right)(-)^{|\nu|}\6_{(\nu)}\big\{\dd{^S}{\phi^i_{(\mu)(\nu)\rho}}
(\vdd{L}{\phi^j}) R^j_\alpha( f^\alpha)\nonumber\\+\vdd{L}{\phi^j}
\dd{^S}{\phi^i_{(\mu)(\nu)\rho}}(R^j_\alpha(
f^\alpha))\big\}\Big][\phi_s(x)](d^{n-1}x)_\rho .\eea In this
equation, a multi-index notation has been used: $(\mu)$ stands for
$\mu_1\dots\mu_k$, $|\mu|$ is the length of the multi-index, i.e.,
if $(\mu)=\mu_1\dots\mu_k$, $|\mu_1\dots\mu_k|=k$ and
$\partial_{(\mu)}=\partial_{\mu_1}\dots\partial_{\mu_k}$.
Furthermore, repeated multi-indices involve sums over both
individual indices and the length. The binomial factors in the
above expression come from the definition of the higher order
Euler operators. The terms in the last line vanish if
$\phi^i_s(x)$ is a one parameter family of solutions to the full
equations of motion, i.e., if equation \eqref{solfull} holds,
whereas the terms in the second line vanish if the (possibly field
dependent) gauge parameters are reducibility parameters of this
one parameter family of solutions, i.e., $f^\alpha=\tilde
f^\alpha$ and equation \eqref{redfull} holds. This finishes the
proof of theorem 1.

In the second part of the appendix, we will show that \bea
I_{\varphi(x)}(S_{\tilde f})[\bar\phi(x)]=
\frac{|\mu|+1}{|\mu|+2}\partial_{(\mu)}\left[
\varphi^j(x)\vdd{}{\phi^j_{(\mu)\rho}}\dd{S_{\tilde f}}{dx^\rho}
\right][\bar\phi(x)]\label{explexfullas} \eea agrees with
\eqref{expllinex} when the latter is evaluated on solutions
$\varphi^i(x)$ of the linearized theory. The only terms that will
contribute to $K_{\tilde f}$ are those for which the derivatives
with respect to the fields of the higher order Euler operators act
on $\partial_{(\lambda)}\vdd{L}{\phi^i}$ contained in $S_{\tilde
f}$ because all other terms will vanish when evaluated on
solutions $\phi^i_s$ of the equations of motion. Taking into
account that \bea L^{\rm free}=\half \frac{\partial^2
L}{\partial\phi^i_{(\sigma)}\partial\phi^j_{(\tau)}}[\bar\phi(x)]
\partial_{(\sigma)}\varphi^i\partial_{(\tau)}\varphi^j,\eea
\eqref{explexfullas} agrees with \eqref{expllinex} evaluated at
$\bar\phi^i(x)$ because \bea
\Big(\dd{^S}{\phi^j_{(\nu)}}(-\partial)_{(\sigma)}\dd{^S
L}{\phi^i_{(\sigma)}}\Big)[\bar\phi(x)]=\dd{^S}{\varphi^j_{(\nu)}}
(-\partial)_{(\sigma)}\Big( \frac{{\partial^S}^2
L}{\partial\phi^i_{(\sigma)}\partial\phi^j_{(\tau)}}[\bar\phi(x)]
\partial_{(\tau)}\varphi^j\Big).\eea Furthermore, if $\phi^i_s(x)$ is a
one parameter family of solutions to the full equations of motion,
$\varphi^i(x)$ is a solution to the linear equations of motion
defined by $L^{\rm free}$. This can be verified by differentiating
\eqref{solfull} with respect to $s$ and putting $s$ to zero.


\begin{thebibliography}{10}

\bibitem{Misner:1970aa}
C.~Misner, K.~Thorne, and J.~Wheeler, {\em Gravitation}.
\newblock W.H. Freeman, New York, 1973.

\bibitem{Abbott:1982ff}
L.~F. Abbott and S.~Deser, ``Stability of gravity with a
cosmological
  constant,'' {\em Nucl. Phys.} {\bf B195} (1982)
76.

\bibitem{Abbott:1982jh}
L.~F. Abbott and S.~Deser, ``Charge definition in nonabelian gauge
theories,''
  {\em Phys. Lett.} {\bf B116} (1982)
259.

\bibitem{ADM}
R.~Arnowitt, S.~Deser, and C.~W. Misner, ``Coordinate invariance
and energy
  expressions in general relativity,'' {\em Phys.\ Rev.} {\bf 122} (1961)
  997--1006.

\bibitem{Regge:1974zd}
T.~Regge and C.~Teitelboim, ``Role of surface integrals in the
{H}amiltonian
  formulation of general relativity,'' {\em Ann. Phys.} {\bf 88} (1974)
286.

\bibitem{Barnich:1995db}
G.~Barnich, F.~Brandt, and M.~Henneaux, ``Local {BRST} cohomology
in the
  antifield formalism. {I}. {G}eneral theorems,'' {\em Commun. Math. Phys.}
  {\bf 174} (1995) 57--92,
\href{http://www.arXiv.org/abs/hep-th/9405109}{{\tt
hep-th/9405109}}.

\bibitem{Barnich:2000zw}
G.~Barnich, F.~Brandt, and M.~Henneaux, ``Local {BRST} cohomology
in gauge
  theories,'' {\em Phys. Rept.} {\bf 338} (2000) 439--569,
\href{http://www.arXiv.org/abs/hep-th/0002245}{{\tt
hep-th/0002245}}.

\bibitem{Anderson:1996sc}
I.~M. Anderson and C.~G. Torre, ``Asymptotic conservation laws in
field
  theory,'' {\em Phys. Rev. Lett.} {\bf 77} (1996) 4109--4113,
\href{http://www.arXiv.org/abs/hep-th/9608008}{{\tt
hep-th/9608008}}.

\bibitem{Torre:1997cd}
C.~G. Torre, ``Local cohomology in field theory with applications
to the
  {E}instein equations,'' \href{http://arXiv.org/abs/hep-th/9706092}{{\tt
  hep-th/9706092}}.
Lectures given at 2nd Mexican School on Gravitation and
Mathematical Physics,
  Tlaxcala, Mexico, 1-7 Dec 1996.

\bibitem{Boulanger:2000rq}
N.~Boulanger, T.~Damour, L.~Gualtieri, and M.~Henneaux,
``Inconsistency of
  interacting, multigraviton theories,'' {\em Nucl. Phys.} {\bf B597} (2001)
  127--171,
\href{http://www.arXiv.org/abs/hep-th/0007220}{{\tt
hep-th/0007220}}.

\bibitem{Barnich:2001jy}
G.~Barnich and F.~Brandt, ``Covariant theory of asymptotic
symmetries,
  conservation laws and central charges,'' {\em Nucl. Phys.} {\bf B633} (2002)
  3--82,
\href{http://arXiv.org/abs/hep-th/0111246}{{\tt hep-th/0111246}}.

\bibitem{LL}
L.~D. Landau and E.~M. Lifshitz, {\em The classical theory of
fields}.
\newblock Pergamon Press, London, 1962.

\bibitem{Bak:1994gf}
D.~Bak, D.~Cangemi, and R.~Jackiw, ``Energy - momentum
conservation in gravity
  theories,'' {\em Phys. Rev.} {\bf D49} (1994)
5173--5181.

\bibitem{Wald:1993nt}
R.~M. Wald, ``Black hole entropy is {N}oether charge,'' {\em Phys.
Rev.} {\bf
  D48} (1993) 3427--3431,
\href{http://arXiv.org/abs/gr-qc/9307038}{{\tt gr-qc/9307038}}.

\bibitem{Iyer:1994ys}
V.~Iyer and R.~M. Wald, ``Some properties of {N}oether charge and
a proposal
  for dynamical black hole entropy,'' {\em Phys. Rev.} {\bf D50} (1994)
  846--864,
\href{http://www.arXiv.org/abs/gr-qc/9403028}{{\tt
gr-qc/9403028}}.

\bibitem{Iyer:1995kg}
V.~Iyer and R.~M. Wald, ``A comparison of {N}oether charge and
{E}uclidean
  methods for computing the entropy of stationary black holes,'' {\em Phys.
  Rev.} {\bf D52} (1995) 4430--4439,
\href{http://arXiv.org/abs/gr-qc/9503052}{{\tt gr-qc/9503052}}.

\bibitem{Jacobson:1994vj}
T.~Jacobson, G.~Kang, and R.~C. Myers, ``On black hole entropy,''
{\em Phys.
  Rev.} {\bf D49} (1994) 6587--6598,
\href{http://www.arXiv.org/abs/gr-qc/9312023}{{\tt
gr-qc/9312023}}.

\bibitem{Henneaux:1992ig}
M.~Henneaux and C.~Teitelboim, {\em Quantization of {G}auge
{S}ystems}.
\newblock Princeton University Press, 1992.

\bibitem{Vinogradov:1977}
A.~Vinogradov, ``On the algebra-geometric foundations of
{L}agrangian field
  theory,'' {\em Sov. Math. Dokl.} {\bf 18} (1977) 1200.

\bibitem{Takens:1979aa}
F.~Takens, ``A global version of the inverse problem to the
calculus of
  variations,'' {\em J. Diff. Geom.} {\bf 14} (1979) 543.

\bibitem{Tulczyjew:1980aa}
W.~Tulczyjew, ``The {E}uler-{L}agrange resolution,'' {\em Lecture
Notes in
  Mathematics} {\bf 836} (1980) 22.

\bibitem{Anderson:1980aa}
I.~Anderson and T.~Duchamp, ``On the existence of global
variational
  principles,'' {\em Amer.\ J.\ Math.} {\bf 102} (1980) 781.

\bibitem{DeWilde:1981aa}
M.~D. Wilde, ``On the local {C}hevalley cohomology of the
dynamical {L}ie
  algebra of a symplectic manifold,'' {\em Lett.\ Math.\ Phys.} {\bf 5} (1981)
  351.

\bibitem{Tsujishita:1982aa}
T.~Tsujishita, ``On variational bicomplexes associated to
differential
  equations,'' {\em Osaka J.\ Math.} {\bf 19} (1982) 311.

\bibitem{Brandt:1990gy}
F.~Brandt, N.~Dragon, and M.~Kreuzer, ``Completeness and
nontriviality of the
  solutions of the consistency conditions,'' {\em Nucl. Phys.} {\bf B332}
  (1990)
224--249.

\bibitem{Dubois-Violette:1991is}
M.~Dubois-Violette, M.~Henneaux, M.~Talon, and C.-M. Viallet,
``Some results on
  local cohomologies in field theory,'' {\em Phys. Lett.} {\bf B267} (1991)
  81--87.

\bibitem{Dickey:1992aa}
L.~Dickey, ``On exactness of the variational bicomplex,'' {\em
Cont.\ Math.}
  {\bf 132} (1992) 307.

\bibitem{Dragon:1996md}
N.~Dragon, ``{BRS} symmetry and cohomology,''
  \href{http://www.arXiv.org/abs/hep-th/9602163}{{\tt hep-th/9602163}}.

\bibitem{Wald:1990aa}
R.~Wald, ``On identically closed forms locally constructed from a
field,'' {\em
  J.\ Math.\ Phys.} {\bf 31} (1990) 2378.

\bibitem{Olver:1993}
P.~Olver, {\em Applications of Lie Groups to Differential
Equations}.
\newblock Spinger Verlag, New York, 2nd~ed., 1993.
\newblock 1st ed., 1986.

\bibitem{Andersonbook}
I.~Anderson,
  ``\href{http://www.math.usu.edu/~fg_mp/Pages/Publications/Publications.html}%
{The variatonal bicomplex},'' tech. rep., Formal Geometry and
Mathematical
  Physics, Department of Mathematics, Utah State University, 1989.
\newblock http://www.math.usu.edu/~fg\_mp/Pages/Publications/Publications.html.

\bibitem{Carter:1968ks}
B.~Carter, ``Hamilton-{J}acobi and {S}chrodinger separable
solutions of
  {E}instein's equations,'' {\em Commun. Math. Phys.} {\bf 10} (1968)
280.

\bibitem{Caldarelli:1999xj}
M.~M. Caldarelli, G.~Cognola, and D.~Klemm, ``Thermodynamics of
  {K}err-{N}ewman-{A}d{S} black holes and conformal field theories,'' {\em
  Class. Quant. Grav.} {\bf 17} (2000) 399--420,
\href{http://www.arXiv.org/abs/hep-th/9908022}{{\tt
hep-th/9908022}}.

\bibitem{Silva:2002jq}
S.~Silva, ``Black hole entropy and thermodynamics from
symmetries,'' {\em
  Class. Quant. Grav.} {\bf 19} (2002) 3947--3962,
\href{http://www.arXiv.org/abs/hep-th/0204179}{{\tt
hep-th/0204179}}.

\bibitem{Bardeen:1973gs}
J.~M. Bardeen, B.~Carter, and S.~W. Hawking, ``The four laws of
black hole
  mechanics,'' {\em Commun. Math. Phys.} {\bf 31} (1973)
161--170.

\bibitem{Carter:1972}
B.~Carter, {\em Black holes}, ch.~Black hole equilibrium states,
pp.~58--214.
\newblock 1972 Les Houches Lectures.
\newblock Gordon and Breach, New York, {C}. and {B}. {D}e{W}itt~ed., 1973.

\bibitem{Townsend:1997ku}
P.~K. Townsend, ``Black holes,''
\href{http://www.arXiv.org/abs/gr-qc/9707012}{{\tt
gr-qc/9707012}}.

\bibitem{Jacobson:1993xs}
T.~Jacobson and R.~C. Myers, ``Black hole entropy and higher
curvature
  interactions,'' {\em Phys. Rev. Lett.} {\bf 70} (1993) 3684--3687,
\href{http://www.arXiv.org/abs/hep-th/9305016}{{\tt
hep-th/9305016}}.

\bibitem{Deser:2002rt}
S.~Deser and B.~Tekin, ``Gravitational energy in quadratic
curvature
  gravities,'' {\em Phys. Rev. Lett.} {\bf 89} (2002) 101101,
\href{http://www.arXiv.org/abs/hep-th/0205318}{{\tt
hep-th/0205318}}.

\bibitem{Deser:2002jk}
S.~Deser and B.~Tekin, ``Energy in generic higher curvature
gravity theories,''
\href{http://www.arXiv.org/abs/hep-th/0212292}{{\tt
hep-th/0212292}}.

\bibitem{Henneaux:2002wm}
M.~Henneaux, C.~Mart\'{\i}nez, R.~Troncoso, and J.~Zanelli,
``Black holes and
  asymptotics of 2+1 gravity coupled to a scalar field,'' {\em Phys. Rev.} {\bf
  D65} (2002) 104007,
\href{http://www.arXiv.org/abs/hep-th/0201170}{{\tt
hep-th/0201170}}.

\bibitem{Martinez:2002ru}
C.~Mart\'{\i}nez, R.~Troncoso, and J.~Zanelli, ``De {S}itter black
hole with a
  conformally coupled scalar field in four dimensions,'' {\em Phys. Rev.} {\bf
  D67} (2003) 024008,
\href{http://www.arXiv.org/abs/hep-th/0205319}{{\tt
hep-th/0205319}}.

\end{thebibliography}

\providecommand{\href}[2]{#2}\begingroup\raggedright\endgroup

\end{document}